\documentclass[conference]{IEEEtran}
\IEEEoverridecommandlockouts
\usepackage{cite}
\usepackage{amsmath,amssymb,amsfonts,amstext}
\usepackage{graphicx}
\usepackage{textcomp}
\usepackage{xcolor}
\def\BibTeX{{\rm B\kern-.05em{\sc i\kern-.025em b}\kern-.08em
    T\kern-.1667em\lower.7ex\hbox{E}\kern-.125emX}}

\usepackage{url}
\usepackage{booktabs}

\usepackage{mathtools}
\usepackage{bm}  % bold math anywhere in math mode and of greek letters too
\usepackage{siunitx}
\usepackage{physics}

\usepackage{ifthen}

\usepackage{multirow}

\DeclareMathOperator*{\argmin}{arg\,min}

\usepackage{algorithm,algpseudocode,float}
\usepackage{setspace} % Needed to fix line spacing inside the algorithms

\algnewcommand{\algorithmicand}{\textbf{ and }}
\algnewcommand{\algorithmicor}{\textbf{ or }}
\algnewcommand{\And}{\algorithmicand}
\algnewcommand{\Or}{\algorithmicor}

\usepackage[firstpageonly=true]{draftwatermark}				

\usepackage{hyperref}
\hypersetup{
    colorlinks=true,
    linkcolor=blue,
    filecolor=magenta,      
    urlcolor=cyan,
    bookmarks=true,
}
\usepackage[noabbrev]{cleveref}  % automatically sort references

\begin{document}

\title{\mbox{}\\ Data-Driven Optimized Tracking Control Heuristic for MIMO Structures: A Balance System Case Study}

\author{Ning Wang, Mohammed Abouheaf and Wail Gueaieb  
	%\thanks{*}% <-this % stops a space
%	\thanks{Mohammed Abouheaf is with College of Energy Engineering, Aswan University, Aswan Egypt and School of Electrical Engineering \& Computer Science, University of Ottawa, Ottawa, Canada. E-mail: mabouhea@uOttawa.ca}%
	\thanks{Ning Wang, Mohammed Abouheaf, and Wail Gueaieb, are with School of Electrical Engineering and Computer Science, University of Ottawa, Ottawa, Ontario, Canada. e-mail: \{nwang094,mabouhea,wgueaieb\}@uOttawa.ca}
}

\maketitle

\DraftwatermarkOptions{%
angle=0,
hpos=0.5\paperwidth,
vpos=0.97\paperheight,
fontsize=0.012\paperwidth,
color={[gray]{0.2}},
text={
  % Disclaimer as in
  % https://journals.ieeeauthorcenter.ieee.org/become-an-ieee-journal-author/publishing-ethics/guidelines-and-policies/post-publication-policies/
  % https://v2.sherpa.ac.uk/romeo/
  \newcommand{\thispaperdoi}{10.1109/SMC42975.2020.9283038}
  \parbox{0.99\textwidth}{This is the postscript version of the published paper. (doi: \href{http://dx.doi.org/\thispaperdoi}{\thispaperdoi})\\
    \copyright 2020 IEEE.  Personal use of this material is permitted.  Permission from IEEE must be obtained for all other uses, in any current or future media, including reprinting/republishing this material for advertising or promotional purposes, creating new collective works, for resale or redistribution to servers or lists, or reuse of any copyrighted component of this work in other works.}},
}

\begin{abstract}
A data-driven computational heuristic is proposed to control MIMO systems without prior knowledge of their dynamics. The heuristic is illustrated on a two-input two-output balance system. It integrates a self-adjusting nonlinear threshold accepting heuristic with a neural network to compromise between the desired transient and steady state characteristics of the system while optimizing a dynamic cost function. The heuristic decides on the control gains of multiple interacting PID control loops. The neural network is trained upon optimizing a weighted-derivative like objective cost function. The performance of the developed mechanism is compared with another controller that employs a combined PID-Riccati approach.  One of the salient features of the proposed control schemes is that they do not require prior knowledge of the system dynamics. However, they depend on a known region of stability for the control gains to be used as a search space by the optimization algorithm. The control mechanism is validated using different optimization criteria which address different design requirements.
\end{abstract}


\section{Introduction}
\label{Introd}
The control problem of balance systems belongs to a class of nonlinear control systems which is usually solved using analytical as well as numerical techniques~\cite{seul_jung_control_2008,nour_fuzzy_2007,yadav_comparative_2011}. The goal of this work is to develop a position-angle regulation scheme of an inverted-pendulum-cart system using two interacting PID control loops. The PID control gains are decided using a nonlinear threshold accepting heuristic. Additional neural network state feedback mechanism is employed to optimize the total dynamic cost during the regulation processes. 
Elmer Sperry introduced a PID scheme in 1911 in order to solve the steering problem of an automatic ship and Nicholas Minorsky designed another type in 1922~\cite{bennett1996brief}. The PID mechanisms are used in applications like manipulator control in robotic arms~\cite{parra2003dynamic}, control of unmanned aerial vehicles~\cite{szafranski2011different}, industrial hydraulic regulators~\cite{rahmat2009application}, temperature control~\cite{shein2012pid}, etc.

Nonlinear Threshold Accepting (NLTA) heuristic is developed by Nahas and Nourelfath, and it relies on a nonlinear accepting threshold criterion formed using a low-pass filter scheme~\cite{nahas2014,Wang_2020}. The NLTA heuristic finds a solution for the optimization problem by continuously updating the local search outcome starting from a random initial feasible guess. The accepting rule employs a magnitude of a low pass filter transfer function. It searches for a better feasible neighboring solution as will be explained later on. This approach is used to solve many NP-hard problems in~\cite{nahas2014}. It tackled energy distribution optimization problems like power system dispatch with prohibition zones and multiple fuel options in~\cite{nahas_non-linear-threshold-accepting_2019}. NLTA is employed to find solutions for the redundancy allocation where it is applied to solve the redundancy allocation problems and enhance the associated reliability in~\cite{nahasnon-linear_2019}. Further, it is employed to regulate the load frequency and automatic voltage disturbances for a network of power generation units in~\cite{Nahas2019}.

The Artificial Neural Network (ANN) is a class of the artificial intelligence sciences, and it is widely used to solve the nonlinear optimization problems. They are employed in power systems to control electric loads where they outperformed other regression approaches~\cite{park1991}. Neural networks are used to implement solutions for a class of adaptive control problems in~\cite{chen1992}. 
An adaptive neural network scheme is employed to design a nonlinear flight controller in~\cite{kim1997}. A data-driven method for computing reachable sets is used to estimate the attractions domains of model predictive controllers in~\cite{Chakrabarty2020}. A dynamic-event triggering control strategy based on integral reinforcement learning is proposed for partially unknown nonlinear systems in~\cite{Mu2020}. In~\cite{abouheaf2019fuzzy}, a fuzzy-neural network approach is employed to control a a flexible wing aircraft.

The work is organized as follows; Section~\ref{probform} explains the dynamical model of a balance system. The development of PID angle-position control loops are detailed out in Section~\ref{TUN}. Further, a nonlinear state feedback control  mechanism is realized using a neural network in Section~\ref{SF}. The usefulness and analysis of the presented schemes are shown in Section~\ref{SIM}. Finally, concluding remarks are pointed out in Section~\ref{CONC}.

%========================================================================

\section{Dynamics of a Balance System}
\label{probform}
The dynamics of a balance system along with the objectives of the control problem are presented in this section. 

\subsection{Inverted-Pendulum-Cart Dynamics}
A free-body-sketch of an inverted-pendulum-cart system is shown in Fig.~\ref{fig:cart}~\cite{prasad_optimal_2014}. The pendulum swings as the moving cart slides along the $x$-direction due to a horizontal force $u$. The masses of a point rigidly connected at the top of the pendulum and the cart are denoted by $m$ and $M$, respectively. The parameters $l$ and $x$ refer to the pendulum's length and displacement of the cart, respectively. The angle $\theta$ is spanned, from the upright reference, by the pendulum. The dynamics of the balance system are arranged as follows~\cite{prasad_optimal_2014}:

\begin{figure}[h]
	\centering
	\includegraphics[width=0.35\textwidth]{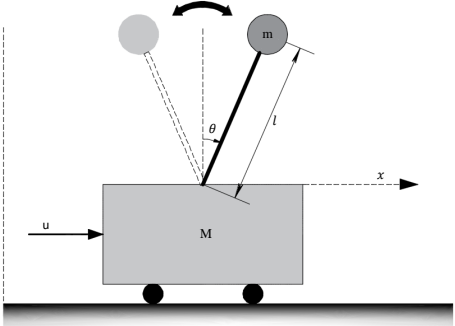}
	\vspace{-10pt}
	\caption{Free-body diagram for a balance system.}
	\label{fig:cart}
\end{figure}

Firstly, the ball-coordinates ($x_p,y_p$) refer to the center of gravity of the inverted pendulum and they are given by
%\begin{equation}
$x_p = x + l\sin\theta$ and $y_p = l\cos\theta$.
%\label{COG}
%\end{equation} 
The force balance in the $x$-direction is expressed as follows
%\begin{equation}
$\displaystyle M\,\frac{d^2\,x}{d\,t^2} \, + \, m\, \frac{d^2\,x_p}{d\,t^2} = u.$
%\label{force1}
%\end{equation} 
%Substituting (\ref{COG}) into (\ref{force1}) yields
Then
%\begin{equation}
$\displaystyle (M + m) \, \ddot{x} \, - \, m \,l \, \sin\theta \,(\dot{\theta})^2 \, + \,  m \, l \, \cos\theta \, \ddot{\theta} \, =\,  u.$
%\label{force2}
%\end{equation} 
Secondly, the force components of the inverted pendulum in the $x$ and $y$ directions are given by
%\begin{equation}
$\displaystyle F_{px} = m \, \frac{d^2\,x_p}{d\,t^2}$ and $\displaystyle  F_{py} = m \, \frac{d^2\,y_p}{d\,t^2}$.
%\label{FC}
%\end{equation} 
The underlying torque equation is given by 
%\begin{equation}
$(F_{px}\cos\theta)\,l - (F_{py}\sin\theta)\,l = (mg\sin\theta)\,l,$
%\label{force3}
%\end{equation}
where $g$ is the gravitational acceleration. Then
%\begin{equation}
$\displaystyle m \, \ddot{x} \, \cos\theta + m \, l \, \ddot{\theta} = m \, g \, \sin\theta.$
%\label{force4}
%\end{equation}
%
%
%Equations (\ref{force2}) and (\ref{force4}) are nonlinear %differential dynamical equations of the balance system so that
%\begin{equation}
%$\displaystyle \ddot{\theta} = \frac{u \, \cos\theta - (M+m) \, g \, \sin\theta  \, + \,  m \, l \, (\cos\theta\sin\theta) \, \dot{\theta}^2}{m \, l \, \cos^2\theta- (M+m) \, l}$
%\label{force5}
%\end{equation}
%and
%\begin{equation}
%$\ddot{x} = \left(u+ m \, l \, \sin\theta \, \dot{\theta^2} - m \, g \, \cos\theta \, \sin\theta\right)/\left(M+m-m \, \cos^2\theta\right).$
%\label{force6}
%\end{equation}
A state space representation can be obtained using means of  Jacobian framework around equilibrium (i.e., $\theta = 0$) so that
\begin{equation}
\dot{X} =  A \, X  \, + \, B\,u,
\label{dyn}
\end{equation}
where
$
X
 =
\begin{bmatrix*}[c]
\theta \\ \dot{\theta} \\ x \\ \dot{x}
\end{bmatrix*}
,
A
 =
\begin{bmatrix*}[c]
0 &  1 &  0 &   0\\
\frac{(M+m)g}{Ml} &  0 &   0 &   0\\
0  &  0  &  0 &   1\\
\frac{-mg}{M}  &  0  &  0  &  0
\end{bmatrix*}
,
B
 =
\begin{bmatrix*}[c]
0\\
\frac{-1}{Ml}\\
0\\
\frac{1}{M}
\end{bmatrix*}
$.

\subsection{Formulation of the Control Problem}
The goal of the optimization or control problem is to let the cart and pendulum follow the desired position-angle trajectories (i.e., $x^{ref}$ and $\theta^{ref}$) using two PID-control loops. 

%========================================================================
\section{Position-Angle Control Mechanism}
\label{TUN}
This section introduces a coupled position-angle PID control mechanism for the balance system using an NLTA approach. 

\subsection{PID Control System}
\begin{figure}[hbt]
	\centering
	\includegraphics[width=0.48\textwidth]{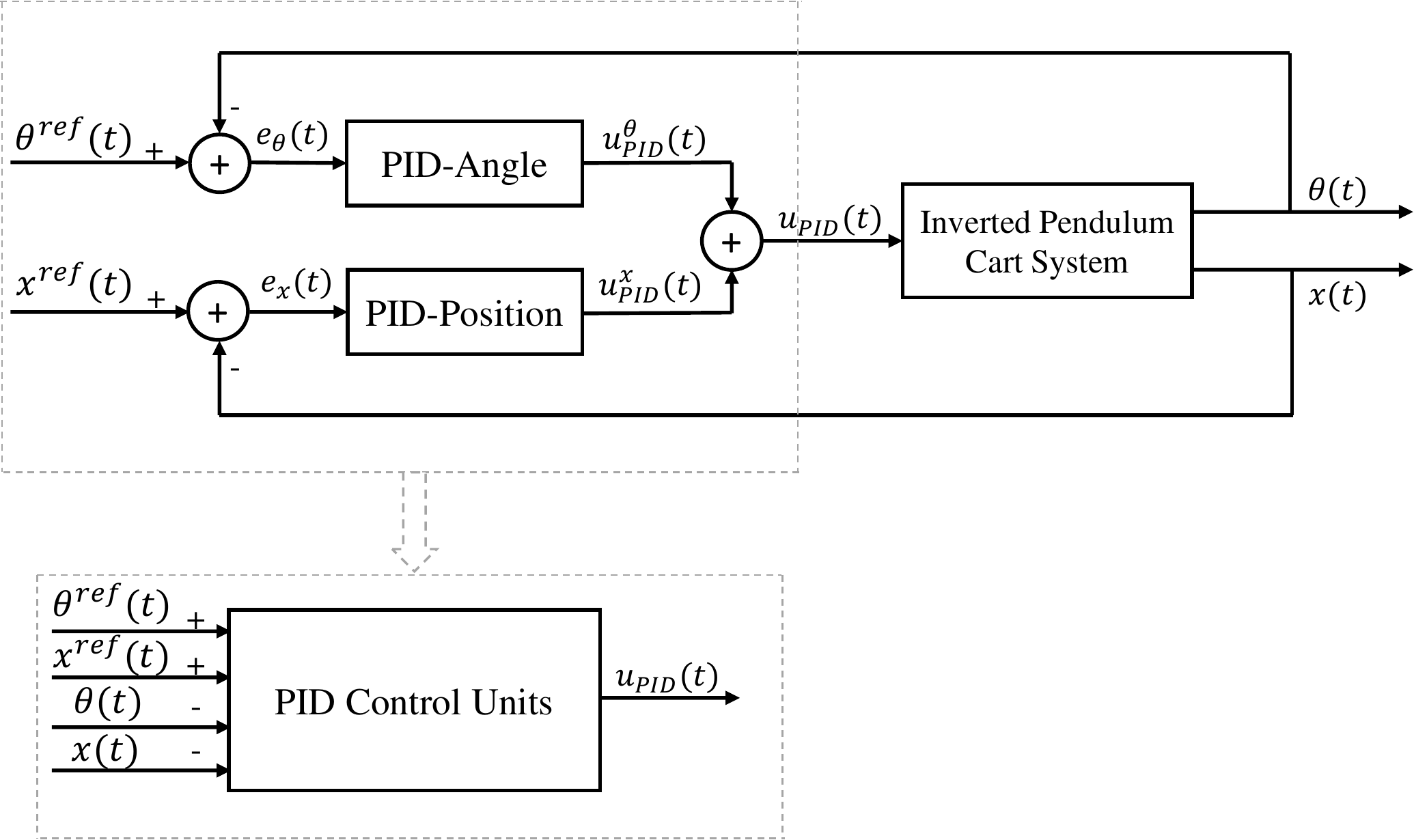}
		\vspace{-10pt}
	\caption{Position-angle PID-control loops.}
	\label{fig:PID}
\end{figure}

%We will now outline the proposed NLTA-based approach for the position-angle control mechanism of the balance system. 
The interacting PID control loops are shown in Fig.~\ref{fig:PID}. It is required to drive the position and angle tracking errors (i.e., $e_x=x^{ref}-x$ and $e_\theta=\theta^{ref}-\theta$) to zeros, respectively.
The underlying control signals $u^{x}_{PID}$ and $u^{\theta}_{PID},$ generated using the PID-control loops, are given by
$ \displaystyle u^{j}_{PID}(t)  = K_p^{j}\,e_{j}(t) + K_i^{j} \int_{0}^{t}\, e_{j}(\eta) \, d\eta + K_d^{j}\, \frac{de_{j}(t)}{dt},j=\{x,\theta\} $ 
%\\
%u^{\theta}_{PID}(t) & = K_p^{\theta}\,e_{\theta}(t) + K_i^{\theta} \int_{0}^{t}\, e_{\theta}(\eta) \, d\eta + K_d^{\theta}\, \frac{de_{\theta}(t)}{dt},  \label{anglePID}
%
where $K^{j}_p$, $K^{j}_i$, $K^{j}_d$ are PID control gains.
The aggregate input control signal is given by $u(t) = u_{PID}(t) = u^{x}_{PID}(t) + u^{\theta}_{PID}(t)$.

As is evident from the dynamics of the balance system, the cart acceleration is coupled to that of the pendulum and vice versa. Therefore, the control gains of the two loops cannot be independently tuned. Instead, both control units are treated by the NLTA heuristic as a single controller with six parameters to tune (i.e., search for the tuple $(K_p^{x},K_i^{x},K_d^{x},K_p^{\theta},K_i^{\theta},K_d^{\theta})$ in a 6-dimensional search space).

\subsection{NLTA-Based PID Gain Tuning}
The NLTA heuristic finds a solution based on a predefined objective function which reflects a specific desired performance. As a matter of fact, one can adopt a cost function to influence the system's transient and steady state characteristics. However, some of these objectives may be contradictory. For example, a shorter settling time may lead to a higher overshoot. Herein, we will suggest a number of cost functions to be adopted by the NLTA approach. Each one exploits a compromise between some of the system's response characteristics.

The first optimization criterion we consider is a convex cost function that minimizes the Integrated Squared Errors (ISE) given by
%\begin{equation}
$\displaystyle  \text{ISE} = \int_{0}^{t}
( w_\theta \, e_{\theta}^2(\eta)  + w_x \, e_{x}^2(\eta) )
d\eta,$
%\label{NLTAobjfunc1}
%\end{equation}
where $w_\theta$ and $w_x$ are some weighting constants. In this case, we took $w_\theta=w_x=0.5$. 
Another objective function is considered to reduce the overshoot along with the ISE. We call it the ``ISE and Absolute Error'' criterion (ISE-AB). It is defined as
%\begin{equation}
$\displaystyle \text{ISE-AB}  =  \int_{0}^{t}
(
w_\theta e_{\theta}^2(\eta)
+ w_x e_{x}^2(\eta)
+ w_\theta' |e_{\theta| (\eta)}
+ w_x' |e_{x}(\eta)|
) d\eta ,$
%\label{PIDobjfuncAB}
%\end{equation}
where $w_\theta$, $w_x$, $w_\theta'$, and $w_x'$, are weight constants, which in this work are initialized to $w_\theta=w_x=w_\theta'=w_x'=0.25$. 
The third objective function tackles the cart response settling time $T_s$ along with the ISE. We call it the ``ISE and Settling Time'' criterion (ISE-TS) and define it by
%\begin{equation}
$\displaystyle \text{ISE-ST} =  \int_{0}^{t}
\left(
w_\theta e_{\theta}^2(\eta)  + w_x e_{x}^2(\eta)
\right)
d\eta \, + \, w_s \, T_s ,
$
%\label{PIDobjfuncST}
%\end{equation}
where $w_s$ is the weight associated to the settling time. Here, the weights are set as $w_\theta=w_x=0.5$ and $w_s=0.1$.
The final objective function addresses the ISE and the cart response overshoot. It is referred to as the ``ISE and OverShoot'' criterion (ISE-OS) and it is expressed as
%\begin{equation}
$\displaystyle 
\text{ISE-OS} = \int_{0}^{t}
\left(
w_\theta e_{\theta}^2(\eta)  + w_x e_{x}^2(\eta)
\right) d\eta \, + \, w_o \, \text{OS} , 
$
%\label{PIDobjfuncOS}
%\end{equation}
%
where $w_o$ is the weight associated to the overshoot. We fixed the weights to $w_\theta=w_x=0.5$ and $w_o=0.1$.
The optimization process using the NLTA approach is detailed out in \Cref{alg:alg4}.
			\begin{algorithm}
	\setstretch{1} % to increase line spacing 
	\caption{NLTA PID-Tuning of Two Control Loops}\label{alg:alg4}
	\begin{algorithmic}[1] % The number tells the line numbering frequency
		\Require
		\Statex $(K_{o,min}^j, \, K_{o,max}^j)$: search ranges of each PID control gain $K_o^j$, $o \in \{p,i,d\}$,  $j\in\{x,\theta\}$.
		\Statex $\Omega_0, \, \Omega^1, \, \Delta \Omega>0$: resonant frequency, initial frequency and a decrement step of the search process, respectively.
		\Statex $N_o$: number of total run times.
		\Statex $N_T$: number of search iterations per each run. 
		\Ensure
		\Statex Tuned gains $K_o^j$, $o \in\{p,i,d\}$,  $j\in\{x,\theta\}$.
		\Statex 
		\For{$q=1$ to $N_o$}
		\State Pick random initial combination of control gains $K_o^j$, $o \in\{p,i,d\}$,  $j\in\{x,\theta\}$ within their feasible ranges. 
		\State Calculate objective function OF by simulating  Fig.~\ref{fig:PID} and computing~(ISE) or any of its variants, e.g.,~(ISE-ST)
		\State OF\_O $\gets$ OF  and  $\Omega \gets \Omega^1$
		%		% 
		\For{$i=1$ to $N_T$}
		\State Select one random gain from its feasible range such that ${K_o^{j}}'$, $o \in\{p, i, d\}$,  $j\in\{x,\theta\}$.
		\label{alg:NLTA:select-candidate-gain}
		\State Use the new neighboring solution and simulate the system to calculate the value of the objective function OF using~(ISE).\Comment{or any of its variants}
		\State OF\_N $\gets$ OF and  $ \displaystyle 
		\norm{H(\Omega)} \gets \frac{1}{\sqrt{1+(\Omega\,/\,\Omega_0)^2}}
		$
		\If{$\displaystyle \frac{\text{OF\_N}}{\text{OF\_O}} \leq 1$ \Or $\displaystyle \frac{\text{OF\_N}}{\text{OF\_O}} \leq \frac{1}{\norm{H(\Omega)}}$}\Comment{The exploration by NLTA is done using the second condition}
		\State OF\_O $\gets$ OF\_N: Interchange the old solution with the better neighboring solution selected using line~\ref{alg:NLTA:select-candidate-gain} 
		\EndIf
		\If{$(\Omega - \Delta\Omega) > 0$}
		\State Adjust the frequency $\Omega \gets \Omega - \Delta\Omega$.
		\EndIf
		\EndFor
		\State OF\_q $\gets$ OF\_O\Comment{lowest cost value for each run}
		\State $\text{tuple}(q) \gets (K_p^\theta,K_i^\theta,K_d^\theta, K_p^x,K_i^x,K_d^x)$ and \quad $\text{OF}(q) \gets \text{OF\_q}$ \Comment{Optimized PID gains for each run  $q$}
		\EndFor
		\State $p \gets \argmin_{q} \text{OF}(q)$
		\State \Return $\text{tuple}(p)$
	\end{algorithmic}
\end{algorithm}

\section{State Feedback Control Mechanism}
\label{SF}
The control interest is not only to regulate the reference-tracking errors but also to optimize a broader objective function that may encompass other signals as well. In the sequel, a state feedback mechanism based on a neural network is developed and then compared to another controller that is based on solving the system's Algebraic Riccati Equation~(ARE)~\cite{prasad_optimal_2014}. 

\subsection{Neural Network Optimization Algorithm}
\begin{figure}[hbt]
	\centering
	\includegraphics[width=0.5\textwidth]{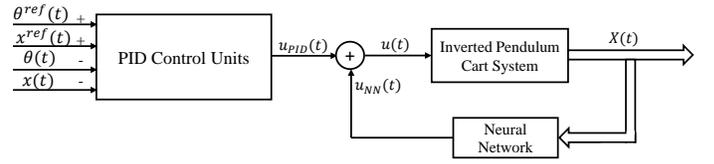}
		\vspace{-20pt}
	\caption{Overall PID-Neural Network control scheme.}
	\label{fig:NN}
\end{figure}

A feedforward NN is trained to optimize a total dynamical cost of the balance system. The overall control scheme including the full state neural network optimization loop and the PID control loops is shown in Fig.~\ref{fig:NN} (i.e., $u=u_{PID}+u_{NN}$). 

The training samples are prepared using a Q-Table process where discretized state-action combinations are employed~\cite{abouheaf2019fuzzy}. Then, according to a performance criteria (i.e., an objective cost function), a suitable control signal is decided. The objective criterion adopted herein is given as follows 
	\begin{equation}
	F_k = X_{k+1}^T\, Q^{NN}\, X_{k+1}+\, R^{NN }\, (u_{k}^{NN})^2,
	\label{NNcost}
	\end{equation}
	where $k$ is a time-index, $Q^{NN} \ge 0 \in \mathbb{R}^{4 \times 4}$ and $R^{NN}>0 \in \mathbb{R}$ are symmetric matrix and positive scalar value respectively. The rest of the details are illustrated in~\Cref{alg:alg3}~\cite{abouheaf2019fuzzy}.
	
\subsection{Linear Quadratic Regulator}

\begin{figure}[hbt]
	\centering
	\includegraphics[width=0.5\textwidth]{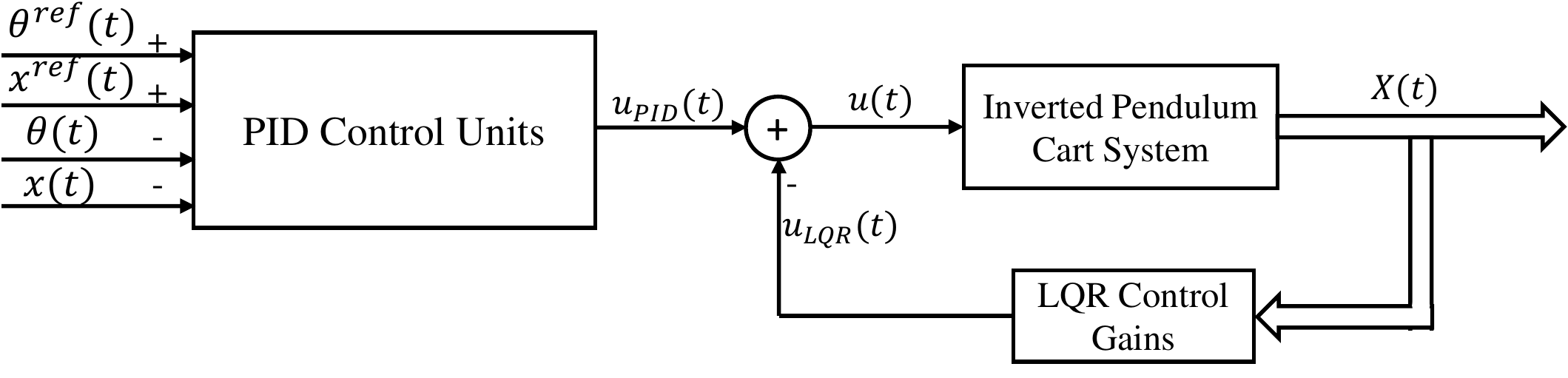}
		\vspace{-20pt}
	\caption{Combined PID-LQR control scheme.}
	\label{fig:lqr}
\end{figure}

The linear quadratic regulator (LQR) approach~\cite{Lewis2012} provides an optimal solution to~\eqref{dyn} while minimizing the quadratic performance index 
$
J = \int_{0}^{\infty}  U(X(\eta),u(\eta)) d\eta,
$
where $U=\frac{1}{2}(X^T(\eta)QX(\eta) +Ru^2(\eta)),$ $Q \ge 0$ and $R > 0$ are weighting matrices, and $X$ is a state vector.
%
%The control performance index is considered to be
%
%\begin{equation}
%J = \int_{0}^{\infty}\, U(X,u)\,dt.
%\label{costFunc}
%\end{equation}
%
The objective is to find the optimal control law $K$ where the optimal control signal is given by $u^o=-K X, \, K = R^{-1}B^TP,$ 
where $P\ge 0\in \mathbb{R}^{n \times n}$ is the solution for the ARE~\cite{Lewis2012}.
%\begin{equation}
%A^T\,P + P\,A - P\,B\, R^{-1}\,B^T\,P + Q = 0.
%\label{ARE}
%\end{equation}
%%
%As such, the closed-loop dynamics can be reformulated as
%%
%\begin{equation*}
%\dot X\, = \,A^c \, X,
%\end{equation*}
%%
%where $A^c=(A \, -B \, R^{-1}\, B^T\, P)$.
%%
The LQR is integrated in the closed loop as shown in Fig.~\ref{fig:lqr}.
	\begin{algorithm}
		%\doublespacing % to increase line spacing in this algorithm
		\setstretch{1} 
		\caption{Neural Network Energy Optimization Scheme}\label{alg:alg3}
		\begin{algorithmic}[1] % The number tells the line numbering frequency
			\Require
			\Statex $(X_{i_{min}},X_{i_{max}}):$ feasible range  of each state $X_i \in X$, $i=1,2,\ldots,n$ with a discrete step $\Delta X_i$.
			\Statex $(u_{min}^{NN},u_{max}^{NN})$ range of $u^{NN}$ with  discrete-step $\Delta u^{NN}$.
			\Ensure
			\Statex Nonlinear State Feedback Neural Network Optimizer NN
			\Statex
			\State \label{alg:NN:rows} The action space $[u_{min}^{NN}, u_{max}^{NN}]$ is discretized 
			into $N_u$ discrete control values: $[u_{1}^{NN}, u_{2}^{NN}, \ldots, u_{N_u}^{NN}]$.
			\State \label{alg:NN:columns}The state space is discretized so that 
			$[
			X_{1_{min}}, X_{1_{min}}+\Delta X_1, X_{1_{min}}+2\Delta X_1, \ldots, X_{1_{max}},
			X_{2_{min}}, X_{2_{min}}+\Delta X_2, X_{2_{min}}+2\Delta X_2, \ldots, X_{2_{max}},
			\ldots,
			X_{n_{min}}, X_{n_{min}}+\Delta X_n, X_{n_{min}}+2\Delta X_n, \ldots, X_{n_{max}}
			]$ and then tuples of all valid combinations of the different states are stored in a single row with $N_X$ entries.
			\State Form a Q-Table $Q(1\ldots N_u, 1\ldots N_X)$, whose row and column indices refer to those of the discrete action and state spaces found as per lines~\ref{alg:NN:rows} and~\ref{alg:NN:columns}, respectively. 
			\State \label{alg:NN:best_strategy} Populate the Q-Table with the objective function~\eqref{NNcost}. Then, decide the control action with the lowest value.
			\State Train a single hidden layer feedforward NN using $N_X$ samples, each with $n$ inputs (combination $X_1, X_2, .., X_n$) and one output (best control decision $u$ as per line~\ref{alg:NN:best_strategy}).
			\State \Return Trained Neural Network $\text{(NN)}$.
		\end{algorithmic}
	\end{algorithm}

\section{Simulation Results}
\label{SIM}
The proposed control schemes are integrated together in a closed-loop structure with the balance system. The system's physical parameters are listed in \Cref{tab:Parameters}~\cite{prasad_optimal_2014}. Hence,
the state space matrices of the system are given by
\begin{align*}
% \label{eq:balance-system-ss-matrices}
& A=
\begin{bmatrix*}[c]    
0 &  1 &  0 &   0\\
29.8615 &  0 &   0 &   0\\
0  &  0  &  0 &   1\\
-0.9401  &  0  &  0  &  0
\end{bmatrix*},
&
& B =
\begin{bmatrix*}[c]
0\\
-1.1574\\
0\\
0.4167
\end{bmatrix*},
&
& X_0 =
\begin{bmatrix*}[c]
0\\
0\\
0\\
0
\end{bmatrix*}.
\end{align*}
\begin{table}[hbt]
	\centering
	\caption{Parameters of the inverted-pendulum-cart system}
	\label{tab:Parameters}
	 \scalebox{0.89}{
	\begin{tabular}{cccc}
		\toprule
		\textbf{Parameter} & \textbf{Value} & \textbf{Parameter} & \textbf{Value} \\
		\midrule
		\multirow{1}{*}{$M$} & $\SI{2.4}{\kg}$ & 
		% \midrule
		\multirow{1}{*}{$m$} & $\SI{0.23}{\kg}$ \\ 
		% \midrule
		\multirow{1}{*}{$l$} & $\SI{0.36}{\m}$ &
		% \midrule
		\multirow{1}{*}{$g$} & $\SI{9.8}{\m/ \s^2}$ \\
		\bottomrule
	\end{tabular}}
\end{table}

The cart and pole reference positions are taken as $x^{ref}(t)=\SI{0.1}{\m}$ and $\theta^{ref}(t)=0$, $\forall t \geq 0$. The simulations are conducted using Matlab-Simulink environment.
%This is to say that the controller is aimed to maintain the pole at an upright position while driving the cart \SI{0.1}{\m} to the right.

\subsection{Performance Analysis of the Different PID Schemes}
At first, the system is simulated with only PID loops, as depicted in Fig.~\ref{fig:PID}. The NLTA algorithm is applied offline to tune the gains of the PID controllers. The parameters adopted for the simulations are listed in \Cref{tab:NLTAP}.

\begin{table}[hbt]
	\centering
	\caption{Parameters of the NLTA optimizer}
	\label{tab:NLTAP}
		 \scalebox{0.9}{
	\begin{tabular}{cccc}
		\toprule
		\textbf{Parameter} & \textbf{Value} & \textbf{Parameter} & \textbf{Value}\\
		\midrule
		\multirow{1}{*}{$(K_{p_{min}}^\theta , K_{p_{max}}^\theta)$} & (-44,-36) & 

		\multirow{1}{*}{$(K_{i_{min}}^\theta , K_{i_{max}}^\theta)$} & (-2,2) \\

		\multirow{1}{*}{$(K_{d_{min}}^\theta , K_{d_{max}}^\theta)$} & (-10,-6) &

		\multirow{1}{*}{$(K_{p_{min}}^x , K_{p_{max}}^x)$} & (-3,1) \\

		\multirow{1}{*}{$(K_{i_{min}}^x , K_{i_{max}}^x)$} & (-2,2) &

		\multirow{1}{*}{$(K_{d_{min}}^x , K_{d_{max}}^x)$} & (-5,-1) \\

		\multirow{1}{*}{$\Omega_0,\Omega_1$ [\si{\rad/\s}]} & $200,50$ & 

		% \midrule
		\multirow{1}{*}{$\Delta \Omega$ [\si{\rad/\s}]} & $0.005$ \\

		\multirow{1}{*}{N\_T} & $1000$  &

		\multirow{1}{*}{N\_o} & $10$ \\

		\bottomrule
	\end{tabular}}
\end{table}

The optimization outcomes, after $10$ search-runs, associated with the objective functions defined earlier are summarized in \Cref{tab:PIDsTunnings}. It is noticed that the specialized objective functions are successful in minimizing their target criteria. For example, the ISE-TS and ISE-OS criteria led to the best settling time and overshoot, respectively.

\begin{table}[h]
	\centering
	\caption{PID tuning outcomes with different cost functions}
	\label{tab:PIDsTunnings}
	 \scalebox{0.9}{
	\begin{tabular}{llcc}
		\toprule
		\textbf{Objective} & \textbf{Optimization} & \textbf{Minimum} & \textbf{Maximum} \\
		\textbf{Function} & \textbf{Criterion} & \textbf{Value} & \textbf{Value} \\
		\midrule
		ISE & Rise Time [\si{\s}] & $0.8122$ & $0.9149$\\ 
		& Settling Time [\si{\s}] & $3.3351$ & $5.0759$\\ 
		& Overshoot [\%] & $1.4904$ & $6.7221$\\
		& ISE & $0.4558$ & $0.4658$\\
		\midrule
		ISE-AB & \multirow{1}{*}{Rise Time} [\si{\s}] & $0.9532$ & $1.3752$\\ 
		&\multirow{1}{*}{Settling Time} [\si{\s}] & $1.8481$ & $3.6533$\\ 
		&\multirow{1}{*}{Overshoot} [\%] & $0.3550$ & $3.1481$\\
		&\multirow{1}{*}{ISE} & $0.4660$ & $0.5347$\\
		&\multirow{1}{*}{ISE-AB} & $3.5746$ & $4.0468$\\
		\midrule
		ISE-ST & \multirow{1}{*}{Rise Time} [\si{\s}] & $1.0120$ & $1.6997$\\ 
		& \multirow{1}{*}{Settling Time} [\si{\s}] & $1.8747$ & $2.8462$\\ 
		& \multirow{1}{*}{Overshoot} [\%] & $1.2320$ & $1.9231$\\
		& \multirow{1}{*}{ISE} & $0.4709$ & $0.5979$\\
		& \multirow{1}{*}{ISE-ST} & $0.6583$ & $0.8825$\\
		\midrule
		ISE-OS & \multirow{1}{*}{Rise Time} [\si{\s}] & $1.1723$ & $1.5896$\\ 
		&\multirow{1}{*}{Settling Time} [\si{\s}] & $3.5311$ & $5.5525$\\ 
		&\multirow{1}{*}{Overshoot} [\%] & $0$ & $0.1948$\\
		&\multirow{1}{*}{ISE} & $0.4735$ & $0.5724$\\
		&\multirow{1}{*}{ISE-OS} & $0.4753$ & $0.5735$\\
		\bottomrule
	\end{tabular}
}
\end{table}

\begin{table}[h]
	\centering
	\caption{PID control gains}
	\label{tab:PIDsGains}
	 \scalebox{0.9}{
	\begin{tabular}{ccccccc}
		% \hline
		\toprule
		\multicolumn{1}{c}{ \multirow{2}*{\textbf{Objective Function}}} & \multicolumn{3}{c}{\textbf{Angle}} \\
		% \cline{2-7}
		\multicolumn{1}{c}{} & \bm{$K_p^\theta$} & \bm{$K_i^\theta$} & \bm{$K_d^\theta$} \\
		\midrule
		PID (Prasad et al (2014)) & -40 & 0 & -8 \\
		% \midrule
		ISE & -43.9238 & 1.2625 & -6.1163 \\
		% \midrule
		ISE-ST & -43.6806 & 0.8948 & -6.2171 \\
		% \midrule
		ISE-OS & -42.3380 & -1.2595 & -6.1730 \\
		% \midrule
		ISE-AB & -43.8129 & 0.2949 & -6.0142 \\
		% \hline
		\midrule
		\multicolumn{1}{c}{ \multirow{2}*{\textbf{Objective Function}}} & \multicolumn{3}{c}{\textbf{Position}} \\
		% \cline{2-7}
		\multicolumn{1}{c}{} & \bm{$K_p^x$} & \bm{$K_i^x$} & \bm{$K_d^x$} \\
		\midrule
		PID (Prasad et al (2014)) & -1 & 0 & -3\\
		% \midrule
		ISE &  -2.8623 & -0.0017 & -3.5402\\
		% \midrule
		ISE-ST &  -2.5071  & -0.0279 & -3.2817\\
		% \midrule
		ISE-OS &   -1.8106 & 0 & -2.6507\\
		% \midrule
		ISE-AB &  -2.3795 & 0 & -3.1028\\
		\bottomrule
	\end{tabular}}
\end{table}

The PID gains obtained using the technique proposed in~\cite{prasad_optimal_2014} and the NLTA approach, using the predefined objective functions, are listed in \Cref{tab:PIDsGains}. 
%The cart and pendulum position responses as well as the control signal and the cumulative cart position error $\int_0^t \abs{e_x(\eta)} \dd{\eta}$ are shown in \Cref{fig:PIDsx,fig:PIDstheta,fig:PIDsu,fig:PIDsaccx}. 
The dynamic cost, transient and steady state characteristics are listed in \Cref{tab:Result1}. 
%
%The results confirm that every objective function successfully achieved its target goal. 
%The best rise time, settling time and overshoot were attained through the ISE, ISE-ST and ISE-OS cost functions, respectively. 
It shows that, the four variants of the PID controllers optimized with the NLTA outperformed the method proposed in~\cite{prasad_optimal_2014} in all aspects.
% and \Cref{fig:PIDsaccx}.
%
Further, the last two columns of \Cref{tab:Result1} assesses performance measures related to the LQR and the ANN optimizers. Although none of these optimizers are enabled yet in the control scheme, they were used as additional measures to compare the results of the various PID tuning schemes.

\begin{table*}[h]
	\centering
	\caption{Performance of PID control units tuned using different cost criteria}
	\label{tab:Result1}
	%\tabcolsep=5pt
	% \textwidth
	\scalebox{0.8}[0.8]{
	\begin{tabular}{ccccccc}
		\toprule
		\textbf{Method} & \textbf{Rise time [\si{\s}]}  & \textbf{Settling time [\si{\s}]} & \textbf{Overshoot [\%]} & \textbf{ISE} & \bm{$\int_{0}^{t}U(X(\eta),u(\eta)) d\eta$} & \bm{$\int_{0}^{t} F(X(\eta),u(\eta)) d\eta$}\\
		\midrule
		\multirow{1}{*}{PID (Prasad et al (2014))} & 4.8914 & 9.6098 & \textbf{0}& 0.72255 & 800.1996 & 79.5381\\ 
		% \midrule
		\multirow{1}{*}{PID (ISE)} & \textbf{0.8900} & 3.3351 & 3.4130 & \textbf{0.4558} & 686.3625 & 67.2464\\
		% \midrule
		\multirow{1}{*}{PID (ISE-ST)} & 1.0120 & \textbf{1.8747} & 1.7828 & 0.4709 & 674.2809 & 66.3176\\
		% \midrule
		\multirow{1}{*}{PID (ISE-OS)} & 1.3283 & 3.8942 & \textbf{0} & 0.5093  & \textbf{662.3581} & \textbf{65.4674}\\
		% \midrule
		\multirow{1}{*}{PID (ISE-AB)} & 1.0578 & 1.9593 & 1.2077 & 0.4751 & 665.5093 & 65.5258 \\
		\bottomrule
	\end{tabular}}
\end{table*}

\subsection{Combined Control Schemes}
\begin{figure}[h]
	\centering
	\includegraphics[width=0.45\textwidth]{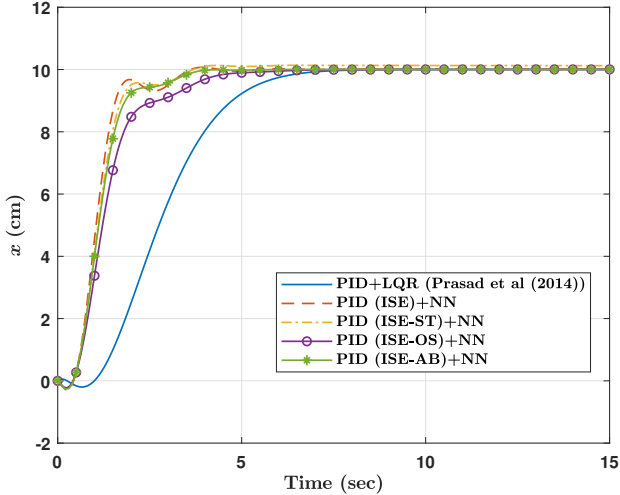}
		\vspace{-10pt}
	\caption{Cart position $x$ (under aggregate control structures)}
	\label{fig:Combinedx}
\end{figure}

\begin{figure}[h]
	\centering
	\includegraphics[width=0.45\textwidth]{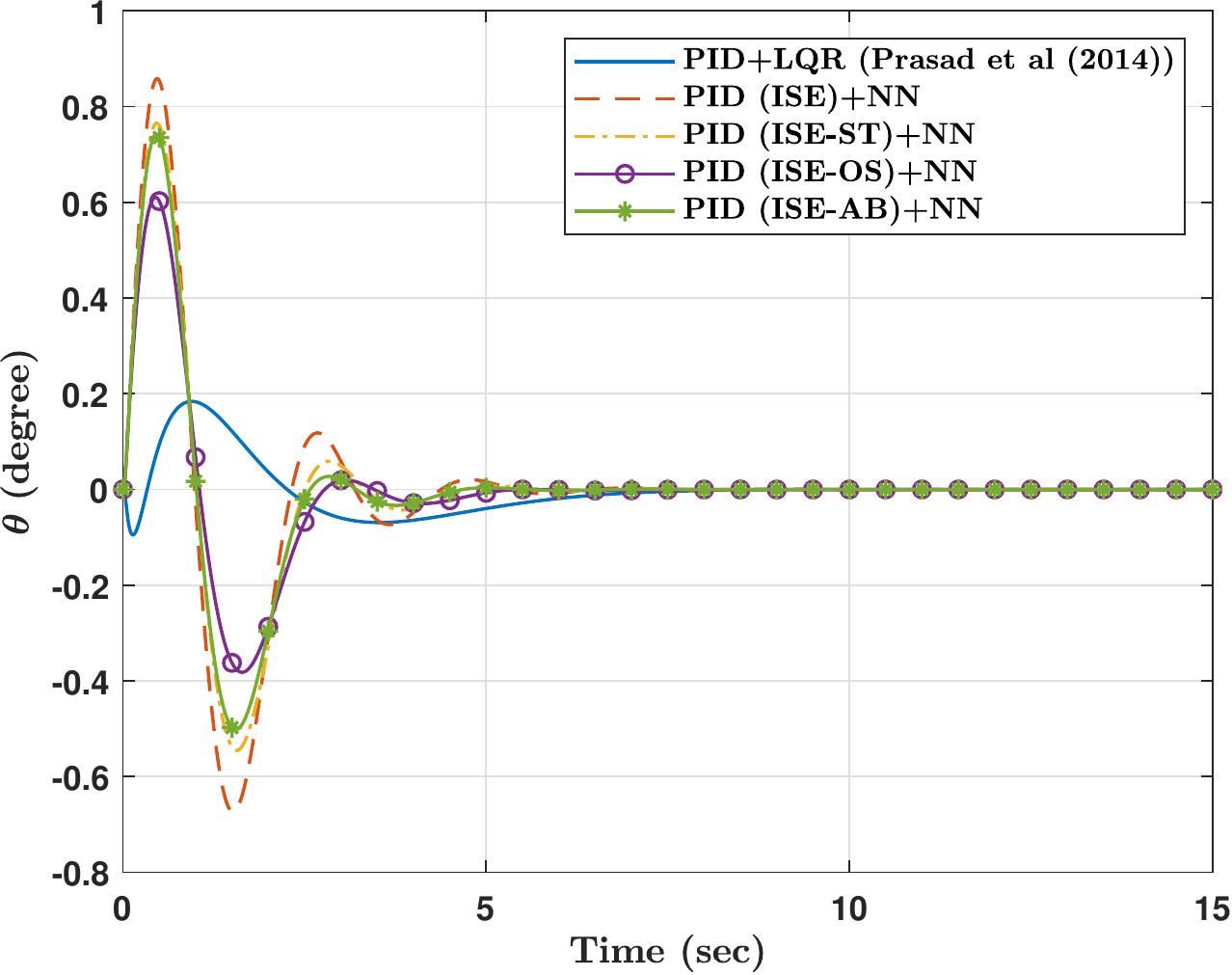}
		\vspace{-10pt}
	\caption{Pendulum angle $\theta$ (under aggregate control structures)}
	\label{fig:Combinedtheta}
\end{figure}

\begin{figure}[h]
	\centering
	\includegraphics[width=0.45\textwidth]{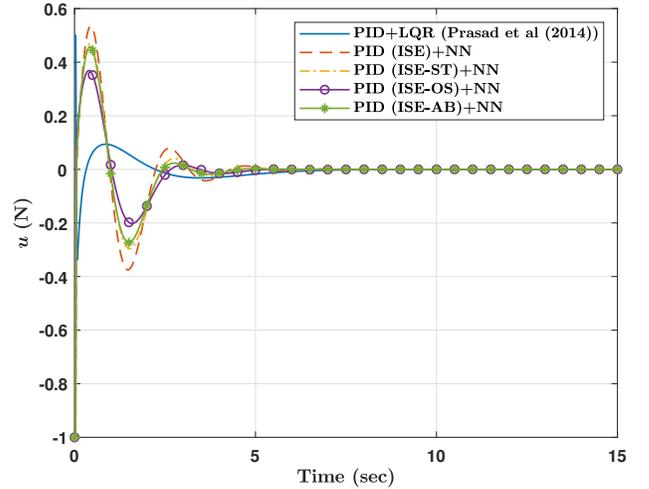}
		\vspace{-10pt}
	\caption{Control signal $u$ (under aggregate control structures)}
	\label{fig:Combinedu}
\end{figure}

The performance of the full aggregate control scheme is studied where the PID and the feedback optimization loops are both enabled. 
The weighting matrices, which are used to find the Riccati solution and to form the objective function of the neural network are, set to
$ Q=\text{diag}\{1,\, 1, \, 500, \, 250\},$
%\begin{bmatrix*}[c]    
%1 &  0 &  0 &   0\\
%0 &  1 &   0 &   0\\
%0 &  0  & 500 &   0\\
%0 &  0  &  0  &  250
%\end{bmatrix*}
$ R = 1,$
$ Q^{NN}=\text{diag}\{1,1,50,25\},$
%\begin{bmatrix*}[c]    
%1 &  0 &  0 &   0\\
%0 &  1 &   0 &   0\\
%0  &  0  &  50 &   0\\
%0  &  0  &  0  &  25
%\end{bmatrix*} 
and $ R^{NN} = 0.016$.
The control law calculated using the Riccati approach is given by
$
K = [ \, -137.7896\,\,-25.9783\,\, -22.3607\,\,-27.5768 \,]
$~\cite{prasad_optimal_2014}.
The ANN training information are detailed in \Cref{tab:NNP}. The cart and pendulum position responses as well as the control signal are shown in Figs.~\ref{fig:Combinedx},~\ref{fig:Combinedtheta},~and~\ref{fig:Combinedu} and the results are summarized in~\Cref{tab:Result2}. Once again, the proposed architecture along with the PID variants outperformed the PID-LQR system suggested in~\cite{prasad_optimal_2014}. The improvement in terms of the measures $\int_0^t U \qty( X(\eta) , u(\eta) ) \dd{\eta}$ and $\int_0^t F \qty( X(\eta) , u(\eta) ) \dd{\eta}$ reached up to 46\% and 47\%, respectively. This reveals how the ANN loop was successful in reducing these two measures with respect to the PID control structure (comparing \Cref{tab:Result1,tab:Result2}). 
\begin{table}[h]
	\centering
	\caption{Parameters of the Neural Network Optimizer}
	\label{tab:NNP}
	\scalebox{0.85}{
		\begin{tabular}{cc}
			\toprule
			\textbf{Parameters} & \textbf{Values} \\
			\midrule
			$X_{min}$, $X_{max}$ & $\pm [\SI{0.175}{\rad} \ \SI{0.35}{\rad/\s} \ \SI{0.1}{m} \ \SI{0.2}{m/s}]$ \\
			$u_{min}$, $u_{max}$ & $\pm \SI{5}{N}$ \\
			$\Delta X$ & $(X_{max} - X_{min})/20$ \\
			$\Delta u^{NN}$ & $(u_{max}^{NN} - u_{min}^{NN})/20$ \\
			Number of hidden layers & $1$ \\ 
			Number of hidden neurons & $13$ \\
			Size of data set & $194,481$ samples \\
			(Training,Validation,Testing)\% & $(70,15,15)\%$ \\ 
%			Validation percentage & $15\%$ \\ 
%			Testing percentage & $15\%$ \\
			Training mechanism & Levenberg-Marquardt \\
			\bottomrule
		\end{tabular}
	}
\end{table} 

\begin{table*}[h]
	\centering
	\caption{Performance using combined control structures (PID+LQR versus PID+NN)}
	\label{tab:Result2}
	\scalebox{0.8}[0.8]{
		\begin{tabular}{ccccccc}
			\toprule
		{\textbf{Method}} & {\textbf{Rise time [\si{\s}}]}  & \textbf{Settling time [\si{\s}]} & \textbf{Overshoot [\%]} & \textbf{ISE} & \bm{$\int_{0}^{t}U(X(\eta),u(\eta)) d\eta$} & \bm{$\int_{0}^{t} F(X(\eta),u(\eta)) d\eta$}\\
			\midrule
			\multirow{1}{*}{PID + LQR (Prasad et al (2014))} & 3.2407 & 6.1969 & \textbf{0} & 1.1437 & 1207.6 & 120.5957\\
			\multirow{1}{*}{PID (ISE) + NN} & \textbf{0.9546} & \textbf{3.3273} & 0.6194 & \textbf{0.4576} & 672.7276 & 65.8731\\
			% \midrule
			\multirow{1}{*}{PID (ISE-ST) + NN} & 1.1275 & 3.5240 & 0.1413 & 0.4733 & 661.1241 & 65.0051 \\
			%	\midrule
			\multirow{1}{*}{PID (ISE-OS) + NN} & 2.0745 & 4.3102 & \textbf{0} & 0.5199 & 659.5133 & 65.1875\\
			% \midrule
			\multirow{1}{*}{PID (ISE-AB) + NN} & 1.2055 & 3.4214 & 0.0103 & 0.4790 & \textbf{654.4692} & \textbf{64.4231} \\
			\bottomrule
	\end{tabular}
}
\end{table*}

\begin{figure}[h]
	\centering
	\includegraphics[width=0.45\textwidth]{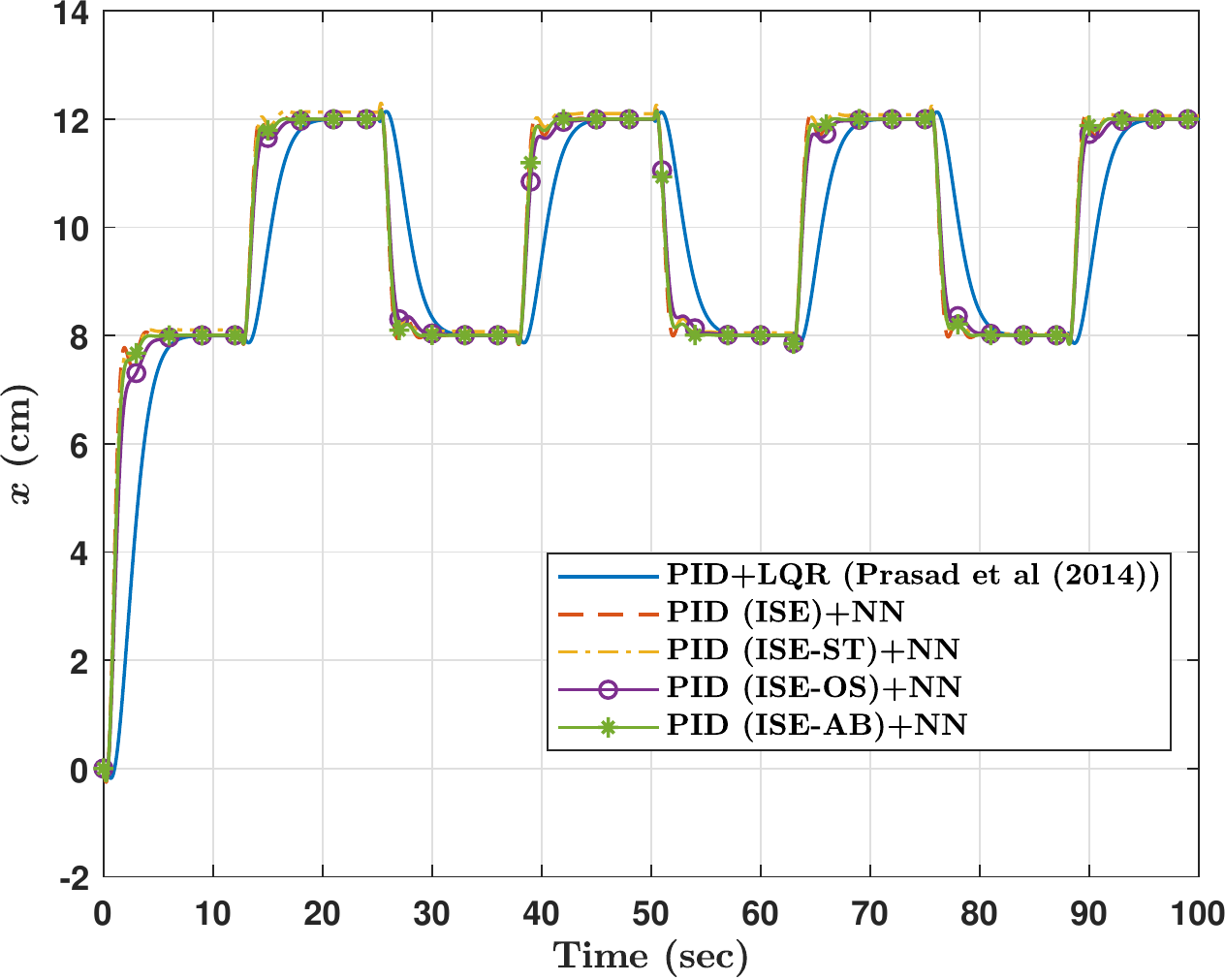}
		\vspace{-10pt}
	\caption{Cart position $x$ (with square wave reference under aggregate control structures)}\label{fig:CombinedxS}
\end{figure}

\begin{figure}[h]
	\centering
	\includegraphics[width=0.45\textwidth]{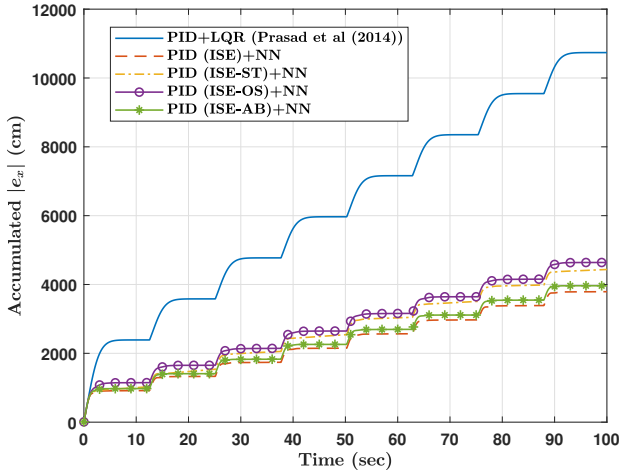}
		\vspace{-10pt}
	\caption{Cumulative cart position error (with square wave reference under aggregate control structures)}\label{fig:CombinedaccxS}
\end{figure}

The cart reference position $x^{ref}(t)$ is assumed to follow a square wave command letting the cart slides back and forth between $8$ and $\SI{12}{\cm}$ using the full control system. The cart position responses and the cumulative cart position tracking-error $\int_0^t \abs{e_x(\eta)} \dd{\eta}$ are displayed in Figs.~\ref{fig:CombinedxS} and~\ref{fig:CombinedaccxS}. The proposed architecture resulted in better transient and steady-state characteristics and better overall dynamic cost than the controller of~\cite{prasad_optimal_2014}.

\section{Conclusion}
\label{CONC}
This work generalizes a nonlinear threshold accepting heuristic along with ANN optimization tool to control multi-output balance systems. The techniques were applied to a system with two outputs and a PID control unit was applied to track each output. It is demonstrated how various objective functions can be integrated in the NLTA scheme to search for optimized PID gains in order to satisfy certain design criteria pertaining to the transient and steady state characteristics. The results were benchmarked against a control algorithm suggested in the literature. It was outperformed by the proposed mechanisms in all conducted simulations with and without the ANN optimization loop. The study showed an improvement in the optimization cost measures of up to 47\%.

\bibliographystyle{IEEEtran}
\bibliography{Bib/mybibliography}

\end{document}